\documentclass[aps,preprint,nofootinbib,eqsecnum]{revtex4}%
\usepackage{amsmath}
\usepackage{graphicx}
\usepackage{amsfonts}
\usepackage{amssymb}%
\providecommand{\U}[1]{\protect\rule{.1in}{.1in}}
\providecommand{\U}[1]{\protect\rule{.1in}{.1in}}

\begin{document}

\begin{center}
{\leftline {USC-07/HEP-B1 \hfill hep-th/0702089}}

{\vskip0.5cm}

{\Large \textbf{Field Theory in 2T-physics with $N=1$ Supersymmetry}%
}\footnote{This work was partially supported by the US Department of Energy,
grant number DE-FG03-84ER40168.}

{\vskip0.5cm}

\textbf{Itzhak Bars and Yueh-Cheng Kuo}

{\vskip0.3cm}

\textsl{Department of Physics and Astronomy, }

\textsl{University of Southern California, Los Angeles, CA 90089-0484, USA.}

{\vskip0.8cm}

\textbf{Abstract}

\end{center}

We construct N=1 supersymmetric field theory in 4+2 dimensions compatible with
the theoretical framework of 2T physics and its gauge symmetries. The fields
are arranged into 4+2 dimensional chiral and vector supermultiplets, and their
interactions are uniquely fixed by SUSY and 2T-physics gauge symmetries. In a
particular gauge the 4+2 theory reduces to ordinary supersymmetric field
theory in 3+1 dimensions without any Kaluza-Klein remnants, but with some
additional constraints in 3+1 dimensions of interesting phenomenological
relevance. This construction is another significant step in the development of
2T-physics as a structure that stands above 1T-physics.

{\vskip0.8cm}

\section{2T-physics Field Theory}

Recently, a field theoretic description of Two-Time Physics (2T physics) has
been established and applied to the Standard Model of Particles and Forces
(SM) \cite{2tstandardM}. The gauge symmetries of 2T-physics can be gauge fixed
such that spacetime and field gauge degrees of freedom are thinned out as one
comes down from 4+2 to 3+1 dimensions holographically without any remnants of
Kaluza-Klein modes.

Many 3+1 spacetimes (different \textquotedblleft times\textquotedblright\ and
\textquotedblleft Hamiltonians\textquotedblright) can be embedded in 4+2
dimensional phase space through the process of this gauge fixing. This
produces dual forms of the 4+2 theory as perceived in 3+1 dimensions. See
Fig.1 in \cite{2tstandardM} for a graphical representation of this property of 2T-physics.

In this letter we will outline the formulation of the general supersymmetric
version of 2T-physics field theory in $4+2$ dimensions, for fields of spins
$0,\frac{1}{2},1,$ with $N=1$ supersymmetry (SUSY). This will be a starting
point for physical applications in the form of the supersymmetric version of
the SM in $4+2$ dimensions, as well as for generalizations to higher $N=2,4,8$
supersymmetric 2T-physics field theory, which will be presented in future papers.

\section{Lagrangian with SUSY}

In what follows, we use mostly left-handed spinors, but also find it
convenient at times to use right handed spinors as the charge conjugates of
left handed ones. The left handed spinor $\psi_{L\alpha}\left(  X\right)  ,$
in the $4$ representation of SO$\left(  4,2\right)  =$SU$\left(  2,2\right)
,$ is labeled with $\alpha=1,2,3,4$\ while the right handed spinor
$\psi_{R\dot{\alpha}}\left(  X\right)  ,$ in the $\bar{4}$ representation of
SU$\left(  2,2\right)  ,$ is labeled with $\dot{\alpha}=1,2,3,4.$ One may also
construct an 8-component spinor of SO$\left(  4,2\right)  $ with a Majorana
condition such that $\psi_{L}$ together with $\psi_{R}$ make up the 8
components of $\psi=\left(
\genfrac{}{}{0pt}{}{\psi_{L}}{\psi_{R}}%
\right)  $ and because of the Majorana condition, $\psi_{R}$ and $\psi_{L}$
are related to each other. One could rewrite all right-handed spinors as
left-handed ones by charge conjugation which is given by
\begin{equation}
\psi_{R}\equiv C\overline{\psi_{L}}^{T}=C\eta^{T}\left(  \psi_{L}\right)
^{\ast},\;\;\text{or\ \ \ }\overline{\psi_{L}}=-\left(  \psi_{R}\right)
^{T}C. \label{cc}%
\end{equation}
Using these definitions we can also write the following equivalent relations
\begin{equation}
\psi_{L}=-C\overline{\psi_{R}}^{T}\text{, \ or\ }\overline{\psi_{R}}=\left(
\psi_{L}\right)  ^{T}C. \label{cc2}%
\end{equation}
Our SO$\left(  4,2\right)  =$SU$\left(  2,2\right)  $ gamma matrices
$\Gamma^{M},\bar{\Gamma}^{M}$ are $4\times4$ matrices in the $4$,$\bar{4}$
Weyl spinor bases, and are given explicitly in footnote (9) of
\cite{2tstandardM}. The antisymmetric charge conjugation matrix is $C=\tau
_{1}\times\sigma_{2}$. The symmetric SU$\left(  2,2\right)  $ metric
$\eta=-i\tau_{1}\times1$ is used to construct the contravariant $\overline
{\psi_{L}}^{\beta}=\left(  \left(  \psi_{L}\right)  ^{\dagger}\eta\right)
^{\beta}=\left(  \psi_{L}^{\ast}\right)  _{\dot{\alpha}}\eta^{\dot{\alpha
}\beta}.$

There is no space here to explain the origin of the 2T-physics gauge
symmetries in field theory that are given in \cite{2tbrst2006}%
\cite{2tstandardM}. But it should be mentioned that it comes from demanding a
local Sp$\left(  2,R\right)  $ symmetry in phase space $\left(  X^{M}\left(
\tau\right)  ,P_{M}\left(  \tau\right)  \right)  $ in the worldline
description of particles. This local symmetry makes position and momentum
indistinguishable at every instant, and requires that the physical space is
the subset of phase space that is Sp$\left(  2,R\right)  $ gauge invariant. We
emphasize the basic important fact that the equations of motion that follow
from the Lagrangian below impose Sp$\left(  2,R\right)  $ gauge singlet
conditions $X^{2}=X\cdot P=P^{2}=0$ in phase space (or OSp$\left(  n|2\right)
$ gauge singlet conditions for a field with spin $n/2$), but now including
interactions \cite{2tstandardM}.

The Sp$\left(  2,R\right)  $ (or OSp$\left(  n|2\right)  $) mentioned above
leads to a corresponding gauge symmetry in the field theoretic formulation of
2T-physics as discussed in \cite{2tstandardM}. To satisfy the gauge symmetries
of 2T-physics, each one of the spin $0,\frac{1}{2},1$ fields and their
interactions can occur only in the form of the Lagrangian terms given below.
One should note that the spacetime structures for kinetic terms, Yukawa
couplings, volume element, etc. are different than usual field theory in 4+2
dimensions. The distinctive space-time features in 4+2 dimensions include the
delta function $\delta\left(  X^{2}\right)  $ and its derivative
$\delta^{\prime}\left(  X^{2}\right)  $ that impose $X^{2}=X^{M}X_{M}=0$, the
kinetic terms of fermions that include the factors $X\bar{D},\bar{X}D,$ and
Yukawa couplings that include the factors $X$ or $\bar{X},$ where
$X\equiv\Gamma^{M}X_{M},$ $\bar{D}=\bar{\Gamma}^{M}D_{M}$ etc$.$ A left arrow
on $\overleftarrow{D}_{M}$ means that the derivative acts on the field on its
left $\overline{\psi_{L}}\overleftarrow{D}_{M}\equiv D_{M}\overline{\psi_{L}}$.

The appearance of explicit factors of $X^{M}$ imply that the action below is
not invariant under translations of $X^{M}$. However, it is invariant under
SO$\left(  4,2\right)  $ rotations of $X^{M}$. This is the right structure for
the theory to have Lorentz symmetry SO$\left(  3,1\right)  $ and translation
symmetry in the 3+1 emergent spacetime $x^{\mu}$ after the gauge fixing of the
2T gauge symmetries. Indeed if we choose the special gauge mentioned at the
beginning of the paper, the emergent spacetime $x^{\mu}$ is Minkowski space,
and in this space SO$\left(  4,2\right)  $ rotations of $X^{M}$ act as the
conformal transformations of $x^{\mu}$ that includes Poincar\'{e} symmetry in
3+1 dimensions.

On the structure demanded by 2T-physics described above we now impose SUSY.
The fields are then organized into chiral supermultiplets $\left(
\varphi,\psi_{L},F\right)  _{i}$ and vector supermultiplets $\left(
A_{M},\lambda_{L},B\right)  ^{a}$ in 4+2 dimensions. These carry indices $a$
for the adjoint representation of a Yang-Mills gauge group $G,$ and indices
$i$ for a collection of representations of the same group. Therefore, all
derivatives will appear as Yang-Mills gauge covariant derivatives $D_{M}$
which take the appropriate form depending on the representation of the group
$G$.

It turns out that the general theory of $N=1$ chiral multiplets coupled to
$N=1$ vector multiplets takes the following form%
\begin{equation}
L=L_{chiral}+L_{vector}+L_{int}+L_{dilaton}. \label{actionsusy}%
\end{equation}
The vector multiplet $\left(  A_{M},\lambda_{L},B\right)  ^{a}$ with its self
interactions is described by%
\begin{equation}
L_{vector}=\delta\left(  X^{2}\right)  \left\{  -\frac{1}{4}F_{MN}^{a}%
F_{a}^{MN}+\frac{i}{2}\left[  \overline{\lambda_{L}}^{a}X\bar{D}\lambda
_{aL}+\overline{\lambda_{L}}^{a}\overleftarrow{D}\bar{X}\lambda_{aL}\right]
+\frac{1}{2}B^{a}B_{a}\right\}  \label{Lvec}%
\end{equation}
The chiral multiplet $\left(  \varphi,\psi_{L},F\right)  _{i}$, with its self
interactions are described by
\begin{align}
L_{chiral}  &  =\delta\left(  X^{2}\right)  \left\{
\begin{array}
[c]{l}%
-D_{M}\varphi^{i\dagger}D^{M}\varphi_{i}+\frac{i}{2}\left(  \overline{\psi
_{L}}^{i}X\bar{D}\psi_{iL}+\overline{\psi_{L}}^{i}\overleftarrow{D}\bar{X}%
\psi_{iL}\right)  +F^{\dagger i}F_{i}\\
+\left[  \frac{\partial W}{\partial\varphi_{i}}F_{i}-\frac{i}{2}\psi
_{iL}\left(  C\bar{X}\right)  \psi_{jL}\frac{\partial^{2}W}{\partial
\varphi_{i}\partial\varphi_{j}}\right]  +h.c.
\end{array}
\right\} \label{Lchi}\\
&  +2~\delta^{\prime}\left(  X^{2}\right)  ~\varphi^{i\dagger}\varphi
_{i}\nonumber
\end{align}
Some of the interactions of the chiral multiplet with the gauge multiplet
already appear through the gauge covariant derivatives $D^{M}\varphi_{i}$ and
$D^{M}\psi_{iL}$. Additional interactions of the vector and chiral multiplets
occur also through the auxiliary and gaugino fields $B^{a}$ and $\lambda
_{L}^{a}$
\begin{equation}
L_{int}=\delta\left(  X^{2}\right)  \left\{  \alpha\varphi^{\dagger i}\left(
t_{a}\right)  _{i}^{~j}\varphi_{j}B^{a}+\beta\varphi^{\dagger i}\left(
t_{a}\right)  _{i}^{~j}\left(  \psi_{jL}\right)  ^{T}\left(  C\bar{X}\right)
\lambda_{L}^{a}\right\}  +~h.c. \label{Lint}%
\end{equation}
where $\alpha,\beta$ will be uniquely determined by SUSY. Finally a sketchy
description of the dilaton is given by%
\begin{equation}
L_{dilaton}=\left\{
\begin{array}
[c]{c}%
-\frac{1}{2}\delta\left(  X^{2}\right)  ~\partial_{M}\Phi\partial^{M}%
\Phi+\delta^{\prime}\left(  X^{2}\right)  \Phi^{2}+\text{superpartners of
}\Phi\\
+\delta\left(  X^{2}\right)  \left\{  \xi_{a}B^{a}\Phi^{2}+V\left(
\Phi,\varphi\right)  \right\}
\end{array}
\right\}  \label{Ldil}%
\end{equation}
We note the following points on the structure of the Lagrangian

$\left(  1\right)  $ The $W\left(  \varphi\right)  $ in $L_{chiral}$ is the
holomorphic superpotential consisting of any combination of $G$-invariant
\textit{cubic} polynomials constructed from the $\varphi_{i}$ (and excludes
the $\varphi^{i\dagger}$)%
\begin{equation}
W\left(  \varphi\right)  =y^{ijk}\varphi_{i}\varphi_{j}\varphi_{k}%
,\;y^{ijk}\text{=constants compatible with }G\text{ symmetry.}
\label{superpotential}%
\end{equation}
The purely cubic form of $W\left(  \varphi\right)  $ leads to a purely quartic
potential energy for the scalars after the auxiliary fields $F_{i}$ and
$B^{a}$ are eliminated through their equations of motion. A purely quartic
potential is required by the 2T gauge symmetry even without SUSY.

$\left(  2\right)  $ The $\bar{X}$ in the Yukawa couplings $\left(  \psi
_{iL}\right)  ^{T}\left(  C\bar{X}\right)  \psi_{jL}\frac{\partial^{2}%
W}{\partial\varphi_{i}\partial\varphi_{j}}$ or $\beta\left(  \varphi^{\dagger
}t^{a}\psi_{L}\right)  ^{T}\left(  C\bar{X}\right)  \lambda_{aL}$ is
consistent with the SU$(2,2)$=SO$\left(  4,2\right)  $ group theory property
$\left(  4\times4\right)  _{antisymmetric}=6$: namely, two left handed
fermions must be coupled to the vector $X^{M}$ to give an SO$\left(
4,2\right)  $ invariant. The $\bar{X}$ insertion is also required for the
2T-gauge invariance \cite{2tstandardM}.

$\left(  3\right)  $ SUSY requires that the dimensionless constants
$\alpha,\beta$ are all determined in terms of the gauge coupling constants $g$
for each subgroup in $G$ as follows\footnote{There is a separate gauge
coupling $g$ for each subgroup in $G,$ so there are separate $\alpha,\beta$
for each $g$.}
\begin{equation}
\alpha=g,\;\beta=\sqrt{2}g,\;
\end{equation}
The only parameters that are not fixed by the symmetries are the Yang-Mills
coupling constants $g,$ and the Yukawa couplings $y^{ijk}$ which are
restricted by $G$-invariance
\begin{equation}
\frac{\partial W}{\partial\varphi_{i}}\left(  t_{a}\varphi\right)  _{i}=0.
\label{Gsymm}%
\end{equation}

$\left(  4\right)  $ Now we turn to the dilaton term $L_{dilaton}.$ As
mentioned above, the superpotential $W\left(  \varphi\right)  $ is restricted
by supersymmetry to be purely cubic in $\varphi$. So for driving the
spontaneous breakdown of the $G$ symmetry the same way as in the
non-supersymmetric case (as in \cite{2tstandardM}), as well as for inducing
soft supersymmetry breaking through the Fayet-Illiopoulos type of term
$\xi_{a}\Phi^{2}B^{a}$, it would be desirable to couple the dilaton $\Phi$ to
the chiral and vector multiplets by having interactions of the form $V\left(
\Phi,\varphi\right)  $ and $\xi_{a}\neq0$ for U$\left(  1\right)  $ gauge
subgroups. However, we have not yet included the superpartners of the dilaton
because this is still under development in the 2T-physics context, so we are
not yet in a position to discuss the SUSY constraints on the desired
couplings. So in this paper we will not be able to comment in detail on the
dilaton-driven electroweak or SUSY phase transition. However, we point out
that in agreement with \cite{2tstandardM} this is again a consistent message
from 2T-physics, namely that the physics of the Standard Model, in particular
the electroweak phase transition that generates mass, is not decoupled from
the physics of the gravitational interactions in a complete unified theory of
all the forces. The full theory may be attained by further pursuing these
hints provided by the 2T-physics formulation of the Standard Model.

\section{SUSY transformations}

We now summarize the properties of the SUSY transformations for the chiral and
vector multiplets that leave invariant the action $S=\int d^{6}xL$. The
dilaton and its superpartners are ignored here. The supersymmetry
transformation for the chiral multiplet is (in the following $\varepsilon
_{R}\equiv C\overline{\varepsilon_{L}}^{T}$ and $\overline{\varepsilon_{R}%
}=\left(  \varepsilon_{L}\right)  ^{T}C,$ and similarly for $\lambda_{R}$ or
$\psi_{R},$ as in Eqs.(\ref{cc},\ref{cc2}))%

\begin{equation}
\delta_{\varepsilon}\varphi_{i}=\left\{  \overline{\varepsilon_{R}}\bar{X}%
\psi_{iL}+X^{2}\left[
\begin{array}
[c]{c}%
-\frac{1}{2}\overline{\varepsilon_{R}}\bar{D}\psi_{iL}+\frac{1}{2}%
\frac{\partial^{2}W^{\ast}}{\partial\varphi^{\dagger i}\partial\varphi
^{\dagger j}}\overline{\psi_{L}}^{j}\varepsilon_{L}\\
-\frac{ig}{2\sqrt{2}}\left(  \overline{\varepsilon_{L}}\lambda_{L}%
^{a}+\overline{\lambda_{L}}^{a}\varepsilon_{L}\right)  \left(  t_{a}%
\varphi\right)  _{i}%
\end{array}
\right]  \right\}  \label{delphi}%
\end{equation}%
\begin{equation}
\delta_{\varepsilon}F_{i}=\overline{\varepsilon_{L}}\left[  X\bar{D}-\left(
X\cdot D+2\right)  \right]  \psi_{iL}-i\sqrt{2}g\left(  \overline
{\varepsilon_{L}}X\lambda_{R}^{a}\right)  \left(  t_{a}\varphi\right)  _{i}.
\end{equation}%
\begin{equation}
\delta_{\varepsilon}\psi_{iL}=i\left(  D_{M}\varphi_{i}\right)  \left(
\Gamma^{M}\varepsilon_{R}\right)  -iF_{i}\varepsilon_{L}\;\;
\end{equation}%
\begin{equation}
\delta_{\varepsilon}\overline{\psi_{L}}^{i}=i\overline{\varepsilon_{R}}%
\bar{\Gamma}^{M}\left(  D_{M}\varphi\right)  ^{\dagger i}+i\overline
{\varepsilon_{L}}F^{\dagger i} \label{delFull}%
\end{equation}
The supersymmetry transformation for the vector multiplet is%

\begin{equation}
\delta_{\varepsilon}A_{M}^{a}=\left\{  -\frac{1}{\sqrt{2}}\overline
{\varepsilon_{L}}\Gamma_{M}\bar{X}\lambda_{L}^{a}+X^{2}\left[
\begin{array}
[c]{c}%
\frac{1}{2\sqrt{2}}~\overline{\varepsilon_{L}}\Gamma_{MN}\left(  D^{N}%
\lambda_{L}^{a}\right) \\
-\frac{ig}{4}\left(  \overline{\varepsilon_{L}}\Gamma_{M}\psi_{R}^{i}\right)
\left(  t^{a}\varphi\right)  _{i}%
\end{array}
\right]  \right\}  +h.c. \label{delAa}%
\end{equation}%
\begin{equation}
\delta_{\varepsilon}B^{a}=\frac{i}{\sqrt{2}}\overline{\varepsilon_{L}}\left[
X\bar{D}-\left(  X\cdot D+2\right)  \right]  \lambda_{L}^{a}+h.c.
\end{equation}%
\begin{equation}
\delta_{\varepsilon}\lambda_{L}^{a}=i\frac{1}{2\sqrt{2}}F_{MN}^{a}\left(
\Gamma^{MN}\varepsilon_{L}\right)  -\frac{1}{\sqrt{2}}B^{a}\varepsilon_{L}%
\end{equation}%
\begin{equation}
\delta_{\varepsilon}\overline{\lambda_{L}}^{a}=i\frac{1}{2\sqrt{2}}\left(
\overline{\varepsilon_{L}}\Gamma^{MN}\right)  F_{MN}^{a}-\frac{1}{\sqrt{2}%
}\overline{\varepsilon_{L}}B \label{delBfull}%
\end{equation}
These SUSY transformations have some parallels to naive SUSY transformations
that one may attempt to write down as a direct generalization from $3+1$ to
$4+2$ dimensions. However, there are many features that are completely
different. These include the insertions that involve $X$ $=X^{M}\Gamma_{M}$ or
$\bar{X}$ $=X^{M}\bar{\Gamma}_{M},$ the terms proportional to $X^{2},$ and the
terms proportional to derivative terms involving $\left(  X\cdot D+2\right)
.$ These are \textit{off-shell} SUSY transformations that include interactions
and leave invariant the off-shell action. The free field limit of our
transformations (i.e. $W=0$ and $g=0$) taken on shell (i.e. terms proportional
to $X^{2}$ and $\left(  X\cdot D+2\right)  $ set to zero) agrees with previous
work which was considered for on-shell free fields without an action principle
\cite{ferrara}.

Despite all of the changes compared to naive SUSY, this SUSY symmetry provides
a representation of the supergroup SU$\left(  2,2|1\right)  $. This is
signaled by the fact that all terms are covariant under the bosonic subgroup
SU$\left(  2,2\right)  ,$ while the complex fermionic parameter $\varepsilon
_{L}$ and its conjugate $\overline{\varepsilon_{L}}$ are in the $4,4^{\ast}$
representations of SU$\left(  2,2\right)  $, as expected for SU$\left(
2,2|1\right)  .$

The closure of these SUSY transformations is discussed in the detailed paper
in the case of the pure chiral multiplet (i.e. gauge coupling $g=0$). The
commutator of two SUSY transformations closes to the bosonic part SU$\left(
2,2\right)  \times$U$\left(  1\right)  \subset$ SU$\left(  2,2|1\right)  $
when the fields are on-shell. More generally, when the fields are off-shell
the closure includes also a U$\left(  1\right)  $ outside of SU$\left(
2,2|1\right)  $ and a 2T-physics gauge transformation, both of which are also
gauge symmetries of the action.

When reduced to $3+1$ dimensions by choosing the special gauge mentioned
earlier, the SU$\left(  2,2|1\right)  $ transformations give a non-linear
off-shell realization of superconformal symmetry in $3+1$ dimensions.

\section{Conserved supercurrent}

The Lagrangian in Eq.(\ref{actionsusy}) transforms into a total divergence
under the SUSY transformations (in the absence of the dilaton). Applying
Noether's theorem we compute the conserved SUSY current. The details are shown
step by step in an upcoming publication. The result is%
\begin{equation}
J_{R}^{M}=\delta\left(  X^{2}\right)  \left\{
\begin{array}
[c]{c}%
D_{K}\left(  X_{N}\varphi^{\dagger i}\right)  \left(  \bar{\Gamma}^{KN}%
\bar{\Gamma}^{M}-\eta^{MN}\bar{\Gamma}^{K}\right)  \psi_{iL}+\frac{\partial
W^{\ast}}{\partial\varphi^{\ast j}}X_{N}\bar{\Gamma}^{MN}\psi_{R}^{j}\\
+\frac{1}{2\sqrt{2}}F_{KL}^{a}X_{N}\left(  \bar{\Gamma}^{KLN}\Gamma^{M}%
-\eta^{NM}\bar{\Gamma}^{KL}\right)  \lambda_{Ra}\\
-\frac{ig}{\sqrt{2}}\varphi^{\dagger i}\left(  t_{a}\varphi\right)  _{i}%
X_{N}\bar{\Gamma}^{MN}\lambda_{Ra}%
\end{array}
\right\}  . \label{JR}%
\end{equation}
The first line comes from $L_{chiral},$ the second from $L_{vector},$ and the
third from $L_{int}.$ The charge conjugate of $J_{R}^{M}$ is the left-handed
counterpart of the above $J_{L}^{M}=-C(\overline{J_{R}^{M}})^{T}$.

To show that this current is conserved we use the equations of motion that
follow from the full Lagrangian. All of the following equations, and their
hermitian conjugates, should be multiplied by $\delta\left(  X^{2}\right)  ,$
so they are required to be satisfied only at $X^{2}=0$%
\begin{equation}
\left(  X\cdot D+1\right)  \varphi_{i}=\left(  X\cdot D+2\right)
F_{i}=\left(  X\cdot D+2\right)  B^{a}=X^{N}F_{NM}^{a}=0
\end{equation}%
\begin{equation}
\left(  X\cdot D+2\right)  \psi_{R}^{i}=\left(  X\cdot D+2\right)  \lambda
_{R}^{a}=0
\end{equation}%
\begin{equation}
D^{2}\varphi^{\dagger i}+\frac{\partial^{2}W}{\partial\varphi_{i}%
\partial\varphi_{j}}F_{j}-\frac{i}{2}\overline{\psi_{Rj}}C\overline{X}%
\psi_{Lk}\frac{\partial^{3}W}{\partial\varphi_{i}\partial\varphi_{j}%
\partial\varphi_{k}}+g\left(  \varphi^{\dagger}B\right)  ^{i}+\sqrt{2}g\left(
\overline{\psi_{L}}t^{a}\right)  ^{i}X\lambda_{R}^{a}=0,
\end{equation}%
\begin{equation}
\left(  D_{M}F^{MN}\right)  ^{a}-if^{abc}\overline{\lambda}_{Lb}\Gamma
^{MN}\lambda_{Lc}X_{M}-ig\varphi^{\dagger}t^{a}\overleftrightarrow{D}%
^{N}\varphi+gX_{M}\overline{\psi_{L}}\Gamma^{MN}t^{a}\psi_{L}=0,
\end{equation}%
\begin{equation}
i\overline{X}D\psi_{R}^{i}+i\overline{X}\psi_{Lj}\frac{\partial^{2}W}%
{\partial\varphi_{i}\partial\varphi_{j}}-\sqrt{2}g\left(  \varphi^{\dagger
}t_{a}\overline{X}\lambda_{L}^{a}\right)  ^{i}=0,
\end{equation}%
\begin{equation}
B^{a}+g\varphi^{\dagger i}\left(  t_{a}\varphi\right)  _{i}=0,\;\;F_{i}%
+\frac{\partial W^{\dagger}}{\partial\varphi^{\dagger i}}=0,
\end{equation}%
\[
i\overline{X}D\lambda_{R}^{a}+\sqrt{2}g\varphi^{\dagger i}\left(
t_{a}\overline{X}\psi_{L}\right)  _{i}=0,
\]
The first two equations impose homogeneity conditions on the fields, while the
others control the dynamics. Using these, and the following crucial Fierz
identities in 4+2 dimensions%
\begin{align}
0  &  =\delta\left(  X^{2}\right)  \frac{\partial^{3}W}{\partial\varphi
_{i}\partial\varphi_{j}\partial\varphi_{k}}\left(  \overline{\psi_{Ri}}\bar
{X}\psi_{Lk}\right)  \left(  \overline{\varepsilon_{R}}\bar{X}\psi
_{Lj}\right)  ,\label{fierz1}\\
0  &  =\delta(X^{2})~f_{abc}\left(  \overline{\varepsilon_{L}}\left[
\Gamma_{M},\bar{X}\right]  \lambda_{L}^{a}\right)  ~\left(  \overline
{\lambda_{L}}^{b}\left[  \Gamma^{M},\bar{X}\right]  \lambda_{L}^{c}\right)  ,
\end{align}
we can verify with some algebra that this SUSY current is conserved
\begin{equation}
\partial_{M}J_{L}^{M}\left(  X\right)  =\partial_{M}J_{R}^{M}\left(  X\right)
=0.
\end{equation}
Besides the direct proof of the invariance of the action given in the detailed
paper, the conservation of the current amounts also to a proof of SUSY for the
theory of Eq.(\ref{actionsusy}) that supplies the equations of motion.

\section{Physical consequences of phenomenological interest}

In a longer paper we will supply the details of this theory and the proof of
supersymmetry in 4+2 dimensions. This construction represents another
significant step in the development of 2T-physics as a structure that stands
above 1T-physics.

After fixing the 2T gauge symmetry in a particular gauge, the 4+2
supersymmetry transformation SU$\left(  2,2|1\right)  $ reduces to the
non-linear superconformal transformation of the corresponding massless fields
in 3+1 dimensions.

The emergent 3+1 SUSY field theory \textit{in this gauge} is in most respects
similar to standard SUSY field theory. However, there are some interesting
additional constraints from the 4+2 structure which would not be present in
the general 3+1 SUSY theory. These may be considered part of the predictions
of 2T-physics. One of these is the banishing of the troublesome
\textit{renormalizable} CP violating terms of the type $\theta\varepsilon
_{\mu\nu\lambda\sigma}Tr\left(  F^{\mu\nu}F^{\lambda\sigma}\right)  .$ This is
good for solving the strong CP violation problem in QCD. This property of the
emergent 3+1 theory already occurs in the non-supersymmetric 2T field theory
as described in \cite{2tstandardM}, and continues to be true also in the
supersymmetric case.

Recalling also that the superpotential cannot have any dimensionful
parameters, we see that phase transitions like supersymmetry breaking and
electroweak breaking need to be driven by the dilaton vacuum expectation
value. Hence according to 2T-physics such phase transitions must be intimately
related to the physics of the supergravity multiplet. One of our immediate
projects is to construct supergravity in the context of 2T-physics and study
its restrictions on the coupling of the dialton to matter. The new
restrictions imposed by 2T-physics would have corresponding phenomenological
consequences which could be of great interest for phenomenological predictions
at the LHC.

Ultimately, the main impact of the 2T-physics point of view is likely to be
along the following ideas. In coming down to 3+1 dimensions there are a
variety of spacetimes that can be obtained through the gauge fixing of the 2T
gauge symmetry, and this is expected to generate a web of dual supersymmetric
field theories. One of the 3+1 images of the 4+2 theory takes the form of the
well-known 3+1 dimensional chiral multiplets coupled to the vector multiplets
as described above. We expect that nonperturbative information can be obtained
from the dual images of this theory. The methods for performing this research
will be discussed in a future paper \cite{ibquelin}.

We gratefully acknowledge discussions with S-H. Chen, B. Orcal, and G. Quelin.

\end{document}